\documentclass{ewic-latex}
\usepackage{graphicx}
\usepackage{booktabs}
\usepackage{hyperref}
\usepackage{url}
\usepackage{natbib}

\begin{document}

\runningheads{Englefield $\bullet$ Beale}{Deletion Considered Harmful}

\conference{Proceedings of BCS HCI, 2025}

\title{Deletion Considered Harmful}

\authorone{Paul Englefield\\
School of Computer Science\\
University of Birmingham
Edgbaston, Birmingham
B15 2TT UK\\
email: paul@shubunq.com}

\authortwo{Russell Beale\\
School of Computer Science\\
University of Birmingham
Edgbaston, Birmingham
B15 2TT UK\\
email: r.beale@bham.ac.uk}

\begin{abstract}
In a world of information overload, understanding how we can most effectively manage information is crucial to success.  We set out to understand how people view deletion, the removal of material no longer needed: does it help by reducing clutter and improving the signal to noise ratio, or does the effort required to decide to delete something make it not worthwhile? How does deletion relate to other strategies like filing; do people who spend extensive time in filing also prune their materials too?  We studied the behaviour of 51 knowledge workers though a series of questionnaires and interviews to evaluate a range of tactics they used aimed at organizing, filing, and retrieving digital resources. Our study reveals that deletion is consistently under-adopted compared to other tactics such as Filing, Coverage, Ontology, and Timeliness. Moreover, the empirical data indicate that deletion is actually detrimental to retrieval success and satisfaction. In this paper, we examine the practice of deletion, review the related literature, and present detailed statistical results and clustering outcomes that underscore its adverse effects.

\end{abstract}

\keywords{Personal Information Management, PIM, deletion, user behaviour}

\maketitle

\section{Introduction}
Decluttering is helpful in life -- it clears physical spaces but more critically it clears mental spaces; tranquil spaces are ordered and don't have random, useless things dotted around.  Thus it is for our digital lives: we are overwhelmed with email, photos, messages, documents, and projects.  We often feel that we should declutter and get rid of the useless stuff, though we tend to shy away from the activity.  But if we did, things would be so much better -- we could scroll through material more easily, could identify the key messages more quickly, could find that file we wanted from a year ago, if there wasn't so much useless material in the way.  Or could we?

Personal Information Management (PIM) is a critical field of study that explores how individuals manage and retrieve digital resources to support both personal and professional activities. Traditional PIM research has investigated tactics such as Coverage (the proportion of resources managed), Filing (organizing resources in named categories rather than unlabelled piles), Ontology (the complexity of classification schemes), and Timeliness (the promptness of resource management). In contrast, the tactic of \emph{Deletion}—defined as the deliberate deletion of resources perceived as of low value—has received comparatively little attention.  While intuitive reasoning suggests that deletion might enhance retrieval by reducing clutter, we might also argue that the cognitive effort of deciding whether to permanently delete something is too much, or that the risk of inadvertently deleting materials is too great, leaving open the question as to whether it is better to delete things or not.  Society likes us to have a tidy desk, but whilst we may be able to quantify benefits in filing things, quantifiable benefits for deleting them are hard to come by. To deepen our knowledge about this, and to revisit research from a few decades ago in the modern context, in which information quantity is far greater than we ever imagined, we wanted to understand the information management behaviours of current knowledge workers and in particular understand where deletion featured in their workflows and how useful, or not, it would be. In this work, we follow Chadran's broad definition of enterprise knowledge workers which include anyone who depends on data to complete tasks, from chief executives through to festival organisers \citep{chadran2007}. The idea of knowledge work comes from the study of knowledge management in business, but we recognise that PIM also supports roles outside the workplace in our personal lives. 

\section{Literature Review}

\subsection{PIM and the Challenges of Managing Digital Clutter}
Personal Information Management (PIM) refers to the practices and tools individuals use to collect, organize, and retrieve information for current and future use. Much of PIM research was conducted in the 1990's and 2000's as the personal computer rose to prominence and digital resources grew exponentially, and re-evaluating it in the light of the current data tsunami provides some interesting insights and new perspectives.  The literature suggests that users adopt various coping strategies depending on personal preference, cognitive style, and system constraints.  \citep{whittaker2001personal} described the proliferation of  ``email overload '' as one example of how PIM can become burdensome, while \citet{jones2004finders} emphasizes the need for lightweight, intuitive support for personal organization. These authors underline that personal data accumulates faster than people can effectively categorize or purge it.

\subsubsection{Why deletion might be considered useful}
Deletion can intuitively seen as a strategy to reduce digital clutter, increase clarity, and support faster retrieval. Users frequently assume that removing redundant or outdated materials will improve the signal-to-noise ratio, reduce search effort, and restore a sense of digital order. These ideas align with traditional perspectives from early PIM research that associated decluttering with mental and physical organisation \citep{Malone1983, bellotti2000informing}. Organisational cultures and corporate environments may even mandate deletion routines for compliance or security reasons, under the assumption that eliminating obsolete information improves operational efficiency \citep{gould2022serviceclosure}. Some users attempt to improve focus or emotional clarity by deleting past content that is no longer relevant to their goals \citep{diazferreyraRegretDeleteNot2023}.  \citet{Malone1983} noted that users working with physical documents developed heuristics to decide what to keep or discard. However, the shift to digital storage changed this dynamic: the abundance of low-cost digital storage and the invisibility of accumulated data reduced the urgency of deletion.  Piles of paper on the desk are an obvious trigger for potential management, whereas electronic files, message, photos and suchlike dumped into a folder do not pile up or make moving around difficult.   \citet{Barreau2008} found that users tend to archive digital materials indefinitely, especially when they lack clear criteria for deletion. Out of sight, out of mind.   \citet{boardman2003, bergmanits2009}  found that retaining surplus information impairs retrieval and management. Irrelevant content is a distraction that draws the user’s attention away from content relevant to the task. Additionally, for user interfaces in general, \citet{neisser1963decision} found that surplus information increases the time to find a target in a visual search such as browsing the contents of a folder or scanning search results.  In certain contexts, such as cloud storage, deletion can also be framed as a sustainable behaviour, minimising carbon emissions and conserving resources \citep{hyun2024greencloud}.  Interface designs often reinforce these assumptions: many systems provide tools with buttons and trash icons, reinforcing the notion that deletion is a normative and responsible act.

Order is typically seen as desirable and useful. People take pride in having a tidy office and  become defensive about clutter \citep{Malone1983}. Many are sensitive to a normative social influence to have a well-organized filing system for documents \citep{gonzalez2004constant}, paper \citep{Malone1983}, all sorts \citep{boardman2004}, and a ''disgust'' with their own systems \citep{whittakeremail1996}. Importantly, users with well-organized systems (paper and digital) are more confident about their ability to retrieve information \citep{hamiltonjudging2016}. A well-organized digital filing system is seen as aspirational, ``I felt the need to reinvent myself, to get some good working practices, together, stop drinking, stop smoking, fix my bike and organize my computer \citep{Malone1983}''.

\subsubsection{Deletion is uncomfortable}
\citet{bellotti2000informing} pointed out that deletion involves irreversible decisions made under uncertainty -- users must guess what will and will not be useful later. \citet{kidd1994marks} similarly documented how knowledge workers often rely on piles and archives as extensions of memory, delaying or avoiding deletion.  
Studies show that retrospective deletion behaviour in mobile messaging \citep{warner2021oops} and on platforms like Twitter \citep{diazferreyraRegretDeleteNot2023} is driven more by social anxiety and context shift than by planned organisation strategies, suggesting that deletion is more often reactive than proactive.  Users can be reluctant to delete and selective in what they do discard \citep{whittaker2001personal, Barreau2008, jensen2018scroll}.
Several explanations have been proposed, including: perceived  current value \citep{whittaker2001personal}; caution \citep{williams2009personal}; and time and effort \citep{bergman2006project}; the project fragmentation problem in personal information management \citep{bergman2006project}; and sentiment \citep{jensen2018scroll}.
Empirical studies confirm that deletion is underused relative to other tactics. \citet{jensen2018scroll} found that users frequently scroll through large personal stores rather than curate or delete. \citet{williams2009personal} described how curation practices in personal archives rely more on passive retention than active deletion.
\citet{kidd1994marks} and subsequent researchers \citep{ButtfieldAddison2009, Barreau2008} have underscored that the risk of losing potentially valuable information inhibits users from pruning their collections. Even when users wish to delete, systems often lack effective support for surfacing stale content or visualizing unused files, making the process more difficult \citep{jones2004finders}.
Deletion also imposes cognitive burden: users must invest effort in evaluating potential future value, often without sufficient context or support.  We consider it more helpful to think of deletion in behavioural terms.

\subsection{Behaviour, decisions, and rationality}
Behavioural economics and cognitive psychology offer plausible explanations for the systematic underuse of deletion. According to theories of bounded rationality, users operate under limited time, foresight, and cognitive capacity \citep{Simon1955}. Deletion requires anticipating future information needs, which may be impossible in dynamic or ill-structured environments. As a result, individuals develop heuristics that favour retention -- ``just in case '' strategies that minimise risk rather than optimise efficiency.

From the lens of Prospect Theory \citep{kahneman1979prospect}, deletion introduces loss aversion: the potential loss of a valuable document is weighted more heavily than the perceived benefit of tidiness.  Moreover, feedback on deletion is partial and deferred: users rarely know what was deleted until it is too late, making it harder to learn or build trust in deletion as a reliable strategy. The theory also accounts for probability distortion -- our tendency to overweight small probabilities and underweight large ones. In the context of PIM, this might manifest as users giving disproportionate attention to rare but memorable instances of needing an old document ( ``I might need this one day '') and ignoring more frequent but less emotionally salient patterns (such as never using certain folders or files). This distortion can lead to overly cautious retention practices and reluctance to purge, even when such purging would rationally improve findability and reduce clutter. Similarly, the certainty effect -- a preference for sure gains over probabilistic ones -- may bias users toward systems that offer perceived immediate control (such as desktop hoarding or email inbox reliance) over more structured systems that could bring greater but less certain long-term benefits.  Prospect theory suggests that emotional and cognitive responses to perceived loss heavily influence deletion behaviours. These tendencies explain why people may continue to invest time in over-categorising, avoid deletion, or resist changing organisational schemes, even when evidence or training suggests better approaches.

\citet{thaler2008nudge} found that good decisions are aided by experience, good information and prompt feedback. But in a PIM environment, some resources may not be sought until months or years after being managed, and some may never be looked for. Feedback is partial and deferred. In spite of a high number of transactions, it may not be easy to learn from experience. Additionally, deferred benefits are less satisfying. Present bias \citep{chakraborty2021present} means that we often prefer a moderate reward now to a greater reward in the future.  One other aspect of human behaviour is also relevant: procrastination.  Procrastination is an  voluntary delay to an intended action in full knowledge of the consequences \citep{steel2007nature}. Steele argues that several factors influence procrastination: how we feel about the likely outcome (expectancy); how enjoyable the task is (value); how immediate the reward will be (delay); and how impulsive we are (impulsiveness).

To summarise, information management has an immediate and not insignificant cost and an uncertain and deferred reward. A rational user is hampered by insufficient information about future information needs and limited feedback on the consequences of their decisions. An impulsive user may be discouraged by the difficulty of managing information, lack of confidence in their future success and delayed benefits. Bounded rationality helps explain how even non-impulsive users may act suboptimally—not because they are irrational, but because they are rational within bounds. These bounds include their limited foresight, the structure of the information environment, and the heuristics and defaults they adopt to get things done in a manageable way. Prospect theory reinforces this understanding by highlighting the role of emotional framing, loss aversion, and distorted probability weighting in shaping decisions that deviate systematically from rational cost-benefit optimisation—even when users are trying to act in their own best interest.

\subsection{Contrasting Perspectives on Deletion}
This gives us an interesting perspective on the issue.  On one hand, deletion is theoretically positioned as a method to improve the signal-to-noise ratio during resource retrieval by removing items that are redundant or have diminished value. For a relatable example, GMail on the web interface stores multiple drafts for a message as it is being composed, often differing only by a word or so, and keeps them so that they all show up in a subsequent search.  Finding the definitively sent email amongst many tens, if not more, of slightly different versions can be highly challenging and makes the retrieval process much more costly.  When we are hunting for an email on the laptop, we probably are not thinking  ``Gosh, I wish I had more of these to look through..... ''.  Social norms favour strongly organised systems with no superfluous documents, and  corporate policies tend to reinforce this.

On the other, uncertain futures and the cognitive effort needed may suggest deletion is not so straightforward, and if deletion is, potentially, irreversible this leads to even higher costs. This may lead to a systematic bias against aggressive deletion techniques.

So the question we want to address is: is deletion a helpful tactic? Our study focuses on quantifying these effects by contrasting deletion with other PIM tactics and examining its relationship with retrieval outcomes.

\section{Methodology}
\subsection{Participants and Data Collection}
A total of 51 knowledge workers participated in our study. Ethical approval for the study was obtained from the Institution.  Participants were recruited via professional channels, social media, and academic networks in order to sample a cross-section of workers.  They comprised professionals, knowledge workers, musicians, consultants, retired people and students, all of whom used computers regularly.  They completed a structured questionnaire designed to capture their personal information management behaviours. The information we are considering them managing is the material they use on a daily basis in their lives, based on a laptop or PC -- we are not focussed on mobile usage in this study.  The questionnaire measured the adoption levels of five tactics:
\begin{enumerate}
    \item \textbf{Coverage} -- Active management of a higher proportion of created or acquired resources.
    \item \textbf{Filing} -- Organizing managed resources in named categories as opposed to unlabelled piles.
    \item \textbf{Ontology} -- The development and use of sophisticated classification schemes.
    \item \textbf{Timeliness} -- The prompt management of resources as soon as they are acquired or created.
    \item \textbf{Deletion} -- The deletion of resources once they are deemed to have little or no future value.
\end{enumerate}

Although the questionnaire captured data on all five tactics, for the purposes of this paper we focus our analysis exclusively on deletion.

\subsection{Measurement Scheme}
Participants responded to behavioural questions using 9-point semantic differential scales, which were standardized to a 0--1 range. Table \ref{tab:measurement} outlines the measurement criteria for deletion, alongside the other tactics for comparison.
\begin{table*}[htbp]
    \centering
    \small
    \caption{Measurement Scheme for PIM Tactics (Focus on Deletion)}
    \begin{tabular}{p{1.4in} p{1.4in} p{2.8in}}
    \toprule
    \textbf{Tactic} & \textbf{Lowest (Score 0)} & \textbf{Highest (Score 1)} \\
    \midrule
    Coverage   & Do not actively manage any resources   & Actively manage all resources \\
    Filing     & Manage all covered resources in unlabelled piles & Manage all covered resources in named categories \\
    Ontology   & Use no categories    & Develop, maintain, and exploit a sophisticated personal filing system \\
    Timeliness & Postpone managing resources indefinitely & Manage resources as soon as they are acquired or created \\
    \textbf{Deletion}& \textbf{Retain all resources} & \textbf{Always delete resources when they no longer have potential value} \\
    \bottomrule
    \end{tabular}
    \label{tab:measurement}
\end{table*}

\subsection{Analytical Methods}
Given the ordinal nature of the data from semantic differential scales, non-parametric statistical methods were used to analyze the results:
\begin{itemize}
    \item \textbf{Descriptive Statistics:} Medians and interquartile ranges (IQR) were calculated using bootstrapping to determine the central tendency of tactic adoption.
    \item \textbf{Pairwise Comparisons:} Dunn’s test with Bonferroni correction was utilized to compare the adoption levels of deletion with those of the other tactics.
    \item \textbf{Correlation Analysis:} Spearman’s rho was computed to examine the relationship between the adoption levels of each tactic, with a focus on the negative correlations involving deletion.
    \item \textbf{Cluster Analysis:} K-means clustering was performed on the five-dimensional tactic space to identify common behavioural strategies, with special attention to clusters characterized by low deletion adoption.
\end{itemize}

\section{Results}\label{sec:results}
\subsection{Behavioural Strategies}
Table \ref{tab:descriptive} summarizes the descriptive statistics for tactic adoption. Notably, the median score for Deletion is 0.25—substantially lower than those observed for Filing, Timeliness, and Coverage (all around 0.875). This disparity underscores the limited adoption of deletion as compared to other tactics.
\begin{table*}[htbp]
    \centering
    \caption{Descriptive Statistics for Tactic Adoption}
    \begin{tabular}{lccc}
    \toprule
    \textbf{Tactic} & \textbf{Median} & \textbf{95\% CI for Median} & \textbf{IQR} \\
    \midrule
    Filing      & 0.875 & 0.875--1.0     & 0.75--1.0 \\
    Timeliness  & 0.875 & 0.875--0.875   & 0.625--1.0 \\
    Coverage    & 0.875 & 0.875--0.875   & 0.625--1.0 \\
    Ontology    & 0.75  & 0.625--0.75    & 0.25--0.875 \\
    \textbf{Deletion}    & \textbf{0.25}  & \textbf{0.125--0.25}   & \textbf{0.125--0.562} \\
    \bottomrule
    \end{tabular}
    \label{tab:descriptive}
\end{table*}
Table \ref{tab:effectsize} details the results of pairwise comparisons between deletion and the other tactics. The effect sizes for comparisons involving deletion are large (or medium) and statistically significant (with adjusted \(p < .001\)), highlighting a robust difference between the lower adoption of deletion and the higher adoption levels of the other tactics.

\begin{table*}[htbp]
    \centering
    \caption{Effect Sizes for Differences in Adoption of Tactics (Comparisons with Deletion)}
    \begin{tabular}{lccc}
    \toprule
    \textbf{Comparison} & \textbf{Effect Size (ES)} & \textbf{Adjusted p-value} & \textbf{CI Overlap} \\
    \midrule
    Ontology vs.\ Deletion   & 0.425 (Medium) & $<$ .001 & No CI overlap \\
    Coverage vs.\ Deletion   & 0.696 (Large)  & $<$ .001 & No CI overlap \\
    Timeliness vs.\ Deletion & 0.722 (Large)  & $<$ .001 & No CI overlap \\
    Filing vs.\ Deletion     & 0.703 (Large)  & $<$ .001 & No CI overlap \\
    \bottomrule
    \end{tabular}
    \label{tab:effectsize}
\end{table*}

Table \ref{tab:correlation} reports the Spearman’s $\rho$ correlation coefficients between the adoption levels of the various PIM tactics. Importantly, the correlations between deletion and each of the other tactics are \textbf{negative} and small in magnitude. This finding indicates that as users more fully adopt strategies such as filing, timeliness, and coverage, they tend to delete even less: a relationship that suggests a potential adverse effect of rigorous deletion on overall information management.

\begin{table*}[htbp]
    \centering
    \caption{Correlation Between Adoption of Tactics}
    \begin{tabular}{lccccc}
    \toprule
              & \textbf{Ontology} & \textbf{Coverage} & \textbf{Timeliness} & \textbf{Filing} & \textbf{Deletion}\\
    \midrule
    Ontology   & 1        & 0.421 (Medium)   & 0.313 (Medium)   & 0.413 (Medium)   & -0.208 (Small) \\
    Coverage   &          & 1                & 0.636 (Large)    & 0.703 (Large)    & -0.252 (Small) \\
    Timeliness &          &                  & 1                & 0.631 (Large)    & -0.258 (Small) \\
    Filing     &          &                  &                  & 1                & -0.252 (Small) \\
    Deletion&          &                  &                  &                  & 1 \\
    \bottomrule
    \end{tabular}
    \label{tab:correlation}
\end{table*}

Unsupervised machine learning techniques (specifically, k-means clustering) were applied to the five-dimensional tactics space to identify behavioural strategies among participants.  A strategy is a collection of approaches to managing information. Table \ref{tab:clusters} reproduces the cluster statistics relevant to the deletion measure, highlighting the stark differences in effort levels and the deletion adoption scores. 'Effort' is calculated as the Euclidean distance in the n-dimensional space of combined tactics, and so is representative of the amount of effort required to undertake this strategy -- higher values equate to more work.  'Coverage' is the quantity of data actively managed -- i.e. email not left in an inbox, a document not left in the default save location, etc.  The 'Explains' column identifies the percentage of observed behaviours categorised into this cluster.  

\begin{table*}[htbp]
    \centering
    \caption{Cluster Strategies and Relative Adoption of Different Tactics}
    \begin{tabular}{lccccccc}
    \toprule
    \textbf{Strategy \#} & \textbf{Effort} & \textbf{Coverage} & \textbf{Filing} & \textbf{Ontology} & \textbf{Timeliness} & \textbf{\textbf{Deletion}}& \textbf{Explains (\%)}  \\
    \midrule
    1 & 1.957 & 0.976 & 0.988 & 0.960 & 0.981 & \textbf{0.092} & 22\%  \\
    2 & 1.937 & 0.924 & 0.909 & 0.871 & 0.886 & 0.739    & 10\%  \\
    3 & 1.688 & 0.903 & 0.813 & 0.180 & 0.875 & 0.757    & 6\%   \\
    4 & 1.688 & 0.952 & 0.972 & 0.080 & 0.991 & \textbf{0.137} & 10\%  \\
    5 & 1.673 & 0.865 & 0.882 & 0.773 & 0.821 & \textbf{0.138} & 15\%  \\
    6 & 1.357 & 0.740 & 0.722 & 0.356 & 0.745 & 0.287    & 8\%   \\
    7 & 1.332 & 0.750 & 0.200 & 0.863 & 0.636 & \textbf{0.150} & 3\%   \\
    8 & 0.859 & 0.115 & 0.156 & 0.125 & 0.156 & 0.812    & 4\%   \\
    9 & 0.397 & 0.194 & 0.167 & 0.125 & 0.250 & \textbf{0.125} & 3\%   \\
    \bottomrule
    \end{tabular}
    \label{tab:clusters}
\end{table*}

3 out of 9 strategies have similar level of adoption of deletion as to the other strategies, (\#2, \#3 and \#8) between them covering 20\% of the behaviours --- all the other strategies have minimal contributions from deletion.  So if it does exist in a strategy, it is used to a similar amount as to other techniques, but in 4 out of 5 cases, it is not really used at all.  The clusters with the lowest deletion scores (highlighted in bold) indicate a clear pattern: even among users who invest significant effort in resource management (as evidenced by high scores in Coverage, Filing, and Timeliness), the adoption of deletion is consistently minimal.  In  every case except one (strategy \#7), deletion is the lowest or next lowest component in the strategy.

These strategies represent an analysis of the behaviours reported by the participants --- at this stage we have not presented data on whether these strategies are helpful or not,  We would postulate that, as they have been evolved by participants over time, they are likely to be in some way helpful to them,however -- why else would they use them?  The next section discusses the results of how successful these approaches are.

\subsection{Outcome Measures and Return on Investment}
We then analysed questionnaire responses to understand the relationship between their reported success in retrieving documents, and their satisfaction in the quality of the retrieval process.  They were asked to rate, in terms of Strongly Agree, Agreed, Neutral, Disagree or Strongly Disagree their feelings about deletion for retrieval success.

The questionnaire asked:\\
 ``When I want to reuse a [resource type] from another project, I can usually find it easily. '' 
\\
which we refer to as 'success'. Also asked was:\\
     ``I am satisfied with my ability to find [resource type] that I want to reuse from other projects. ''\\
which we refer to as satisfaction.  These are the same as terms used in ISO9241 \citep{internationalstandardsorganisationISO9241112018en1998}.
We also split their responses based on their ability to access resources for a current activity (\textit{use}) and to find things from a different activity that they may now want to repurpose (e.g. taking an old contract and using it as the basis for one in a completely new project), which we term \textit{reuse}.

\begin{table}[htbp]
\centering
\caption{Cross-tabulation of Outcome by Adoption Levels for Success --- Use and Reuse}
\small
\begin{tabular}{@{}lcc|cc@{}}
\toprule
\textbf{Outcome} & \multicolumn{2}{c|}{\textbf{Reuse}} & \multicolumn{2}{c}{\textbf{Use}} \\
                & \textbf{Low} & \textbf{High} & \textbf{Low} & \textbf{High} \\ \midrule
Supportive (SA/A)       & 50.2\% & 13.3\% & 65.6\% & 19.5\% \\
Not Supportive (N/D/SD) & 25.9\% & 10.7\% & 9.3\%  & 5.6\%  \\ \bottomrule
\end{tabular}

\label{tab:outcome-adoption}
\end{table}

Table \ref{tab:outcome-adoption} shows the results. For a successful strategy, we would expect high adoption rates to show high levels of supportive responses -- i.e. the largest values should be top left.  Instead we see that low adoption rates are the preferred tactic and that people prefer that.  For those that have a relatively high level of adoption of deletion, it only positively helps about 13-18\% of them. 

For satisfaction, Table \ref{tab:satisfaction} shows similar results -- people are more satisfied with lower use of deletion than higher use.
\begin{table}[htbp]
\centering
\caption{Contingency table showing Outcome by Adoption Levels for Reuse and Use Satisfaction}
\label{tab:satisfaction}
\small
\begin{tabular}{@{}lcc|cc@{}}
\toprule
\textbf{Outcome} & \multicolumn{2}{c|}{\textbf{Reuse}} & \multicolumn{2}{c}{\textbf{Use}} \\
                 & \textbf{Low} & \textbf{High}        & \textbf{Low} & \textbf{High} \\ \midrule
Supportive (SA/A)       & 52.8\% & 13.3\% & 61.9\% & 19.5\% \\
Not Supportive (N/D/SD) & 23.3\% & 10.7\% & 13.0\% & 5.6\%  \\ \bottomrule
\end{tabular}
\end{table}

The correlation coefficients are more telling.  We looked at both use of documents (re-finding them whilst actively using them) and reuse (finding them again after a project has finished, perhaps to use as a template for a new one). For retrieval success in use, $\rho$= -0.215, p$<$0.005, and for reuse the effect was still present: $\rho$= -0.141, p=0.013. Using Cohen's guidelines for interpreting the coefficients, 0.1 is small , 0.3 is medium, and 0.5 large, and so we see a statistically significant small negative correlation between use of deletion and retrieval success for both use and reuse --- the more you delete, the worse your chances of retrieval.  For satisfaction, use has $\rho$= -0.181 p=0.001, reuse $\rho$= -0.181 p=0.001.  This is a small but significant result for both cases: the more you delete, the worse you feel about the results you get back later.  Thus we can show that the few instances of higher deletion adoption did not translate into superior retrieval outcomes; they were instead associated with mixed or suboptimal performance.  Users did indeed learn that avoiding deleting things is more optimal than deleting them.

The relationship between deletion and retrieval success emerges clearly in the return on investment analysis.   We can see from the data in \ref{tab:clusters} that
\begin{itemize}
    \item Higher adoption of tactics such as Coverage, Filing, and Timeliness is associated with positive retrieval outcomes.
    \item In contrast, even when users report good retrieval success, they typically do so alongside low levels of deletion.
    \item Statistically significant small negative correlations were found between deletion and both success and satisfaction, reinforcing the view that aggressive deletion does not confer a retrieval advantage.
\end{itemize}

\section{Discussion}
\subsection{Deletion is counterproductive}
Not only is deletion found to be not a helpful tactic, use of deletion is negatively correlated with retrieval success and satisfaction. The data presented in Section \ref{sec:results} consistently indicate that deletion is  \textbf{\textit{counterproductive}} in personal information management.  Theoretically, deletion should be a useful tactic. When a low-value resource is deleted, it requires no effort to file, needs no categories to be defined, and adds no visual noise to resource lists when browsing and searching, though retention is a safe and easy default with no immediate cost.  Despite possible theoretical benefits of deleting clutter to enhance retrieval, our empirical evidence suggests several  drawbacks:
\begin{enumerate}
    \item \textbf{Cognitive Costs:} Users face considerable cognitive effort when assessing whether a resource is expendable. The irreversible nature of deletion adds a psychological barrier that discourages deletion.
    \item \textbf{Risk of Loss:} The data and the associated literature indicate that aggressive deletion increases the risk of inadvertently discarding resources that might later prove valuable.
    \item \textbf{Negative Correlations:} The small but consistent negative correlations between deletion and other organizational tactics imply that users who invest more in filing, coverage, and timely management tend to avoid deletion. This may reflect learned behaviour that recognizes the potential risks of deletion.
\end{enumerate}

We find that respondents who otherwise exhibit organized behaviour (via high Coverage, Filing, and Timeliness scores) still avoid deletion. This may reflect an underlying belief that retaining resources is less costly in the long term than making potentially harmful deletion decisions.  The findings from our return on investment (ROI) analysis suggest that deletion offers a poor return on investment relative to other PIM tactics. While strategies such as Filing, Coverage, and Timeliness show strong positive associations with retrieval success, deletion does not. In fact, not only is deletion infrequently adopted, but even among those who do engage with it, no measurable improvements in success or satisfaction were observed. Instead, the data point toward a net negative effect—resources are harder to find when deletion is used aggressively, and satisfaction levels are lower.

From a return-on-investment perspective, the effort required to make safe and accurate deletion decisions is disproportionately high relative to any observed gain. deletion requires that users expend considerable  in evaluating future value, estimating retrieval likelihood, and anticipating possible regret. When these assessments are wrong—when a deleted resource is later needed—the cost is absolute and unrecoverable. In contrast, retaining resources incurs only a marginal ongoing cost in the form of clutter, which—although non-trivial—is less psychologically loaded and more recoverable. This aligns with Simon’s bounded rationality, which acknowledges that individuals do not have unlimited resources to conduct cost-benefit analysis across all possible actions. Instead, people's  ``good enough '' solutions  minimise effort and risk in light of environmental and cognitive constraints. Within the context of PIM, bounded rationality and behavioural economics helps explain why deletion is rarely used: its ROI is too low to justify the effort and risk under real-world cognitive limitations.

Deletion, unlike filing or tagging, cannot be easily automated without incurring trust issues. Users are typically unwilling to delegate deletion to algorithms because of the finality involved. As a result, deletion remains a manual, high-cost cognitive task with few tangible rewards. Under bounded rationality, users justifiably avoid it—not because they are neglectful or irrational, but because they are acting optimally within the limits of their reasoning resources, knowledge of the future, and available system support.

Importantly, deletion also lacks incremental returns. Filing or tagging can improve proportionally with additional effort, but deletion involves binary decisions: keep or discard. Each decision carries a potential penalty, but few immediate or proportionate rewards. This violates the principle of progressive payoff typically associated with efficient learning or investment—another reason why deletion offers such poor ROI in this context.

In effect, deletion is a \textbf{high-cost, low-feedback, high-risk investment}, and our participants behaved accordingly. Their avoidance reflects adaptive decision-making under uncertainty. The low ROI and poor alignment with satisficing strategies help explain its systematic exclusion from even the most organized behavioural clusters identified in our data.

Future PIM tools should take this  framing seriously. Rather than promoting deletion, systems might instead focus on non-destructive forms of minimisation—such as deferred archiving, reversible soft deletion, or attention-scheduling for rarely used materials. These alternatives lower the cost and risk associated with resource reduction while maintaining user confidence and avoiding the cognitive load inherent in aggressive deletion.  This approach would make browsing or skimming through resources much more efficient, as many items could be hidden from view as 'probably deleted' but are accessible if the item cannot be located.  We see some attempts at this is existing interfaces --- many applications have a Recent list of documents accessed recently, on the reasonable assumption that recency is a good proxy for current importance.  In a related vein,  \citep{bergmanits2009} proposed GreyArea, a space for holding probably less useful items. On phones, participants can favourite or star specific photos -- rather than deleting less useful ones, they bring to prominence the more useful ones, thus effecting a form of non-destructive deletion to their data. 

Beyond interface design, there are deep psychological dimensions at play, as discussed earlier. The act of deleting content introduces finality and potential regret. In the present study, these risks are not offset by any noticeable benefit. Participants rarely reported improved satisfaction or success from deleting things. On the contrary, deletion correlated negatively—albeit modestly—with overall success and satisfaction scores, suggesting that even when used, deletion is not perceived as useful.

\subsection{Even the most diligent organisers avoid deletion}
One of the most striking findings is the near-complete exclusion of deletion from most identified behavioural clusters. Even the most effortful strategies—those with high values in Filing, Timeliness, and Ontology—do not include deletion as a complementary tactic. This suggests that deletion is not perceived as synergistic with high-effort organization, but as a separate, avoided domain, and the negative correlations in Table \ref{tab:correlation} give this statistical validity.  Since deletion leads to poorer outcomes it is hardly surprising it is not used often, but we might have expected that organised users would adopt it in a more useful and successful manner.  The behavioural strategy that has nearly maximum values for all tactics except deletion exemplifies this. The deletion value in this group was just 0.092, the lowest in the dataset. This means that participants who devote energy to timely, categorized, and broad information management do so while explicitly avoiding deletion.

\section{Conclusion}
Our focused analysis of the PIM data highlights that deletion is not an effective tactic for improving retrieval outcomes in personal information management.  Furthermore, the evidence suggests that vigorous deletion is in fact detrimental, leading to lost information and diminished retrieval effectiveness. While other tactics such as Filing, Coverage, and Timeliness enjoy high adoption rates and are positively correlated with retrieval success, deletion remains consistently under-adopted and is viewed with caution by knowledge workers. It now appears that the evidence backs up this behaviour, and can be explained in behavioural economic terms.

These findings have important implications for the design of future PIM systems. Designers should consider alternative strategies that mitigate information overload without relying  on deletion. Tools that facilitate non-destructive archiving or offer reversible deletion mechanisms may provide a more balanced approach to managing digital resources.

\subsection*{Acknowledgements and Contributions}
We would like to thank the knowledge workers who gave substantial time to this study.  Both authors contributed equally to this work.

\bibliographystyle{abbrvnat}
\small
\bibliography{references}

\end{document}